\documentclass[sigconf]{acmart}
\AtBeginDocument{}

\setcopyright{acmcopyright}
\copyrightyear{2024}
\acmYear{2024}

\acmDOI{XXXXXXX.XXXXXXX}

\acmConference[CIKM `24]{CIKM 2024: Conference on Information and Knowledge Management}{October 21 – 25, 2024}{Boise, ID}

\usepackage[utf8]{inputenc} 
\usepackage[T1]{fontenc}    
\usepackage{url}            
\usepackage{booktabs}       
\usepackage{amsfonts}       
\usepackage{nicefrac}       
\usepackage{microtype}      
\usepackage{xcolor}         
\usepackage{graphicx}

\usepackage[colorinlistoftodos]{todonotes}
\usepackage{xspace}

\usepackage{multirow}
\usepackage{multicol}
\usepackage{makecell}
\usepackage{subfig}
\usepackage{slashbox}

\usepackage{acronym}
\usepackage{arydshln}
\usepackage{booktabs}
\usepackage{wrapfig}

\usepackage{mathrsfs}
\usepackage{mathtools}
\usepackage[inline]{enumitem}
\usepackage{xcolor, soul}
\sethlcolor{yellow}
\usepackage{tikz}
\usepackage{listings}
\usepackage[frozencache,cachedir=.]{minted}
\usetikzlibrary{shapes.misc, positioning}

\usepackage{balance}
\usepackage{amsmath}

\newcommand{\ift}{Mistral-7B-Instruct-v0.2 }
 
\newcommand{\nileift}{Generic-IFT}
\newcommand{\nileevolve}{Generic-IFT-Evolve}
\newcommand{\reaperift}{REAPER-IFT}

\title{REAPER: Reasoning based Retrieval Planning for Complex RAG Systems}

\author{Ashutosh Joshi*, Sheikh Muhammad Sarwar*, Samarth Varshney*, }
\author{Sreyashi Nag, Shrivats Agrawal, and Juhi Naik}
\thanks{*these authors contributed equally}
\email{(jashutos, smsarwar, varshsam, sreyanag, shrivagr, juhinaik) @amazon.com}

\begin{document}

\begin{abstract}

Complex dialog systems often use retrieved evidence to facilitate factual responses. Such RAG (Retrieval Augmented Generation) systems retrieve from massive heterogeneous data stores that are usually architected as multiple indexes or APIs instead of a single monolithic source. For a given query, relevant evidence needs to be retrieved from one or a small subset of possible retrieval sources. Complex queries can even require multi-step retrieval. For example, a conversational agent on a retail site answering customer questions about past orders will need to retrieve the appropriate customer order first and then the evidence relevant to the customer's question in the context of the ordered product. Most RAG Agents handle such Chain-of-Thought (CoT) tasks by interleaving reasoning and retrieval steps. However, each reasoning step directly adds to the latency of the system. For large models this latency cost is significant -- in the order of multiple seconds. Multi-agent systems may classify the query to a single Agent associated with a retrieval source, though this means that a (small) classification model dictates the performance of a large language model.  In this work we present REAPER (\textbf{REA}soning-based \textbf{P}lann\textbf{ER}) - an LLM based planner to generate retrieval plans in conversational systems. We show significant gains in latency over Agent-based systems and are able to scale easily to new and unseen use cases as compared to classification-based planning. Though our method can be applied to any RAG system, we show our results in the context of a conversational shopping assistant.

\end{abstract}

\maketitle

\section{Introduction}
Conversational shopping assistants help customers navigate their shopping journey by providing relevant information at the right time. They are equipped to answer customer questions on shopping needs, products, comparisons, make recommendations based on this context, and facilitate product discovery. A conversational shopping assistants is thus trained on both product catalog and open data sources. It uses a RAG (Retrieval Augmented Generation) framework~\cite{NEURIPS2020_6b493230} where the response to a customer's query is generated by an LLM, using evidence from one or more retrieval sources. Most complex dialog systems cover a large variety of topics. They need to retrieve evidence from data stores and indexes that are potentially petabytes in size and store heterogeneous documents in multiple modalities. These massive data stores are usually structured as multiple homogeneous indexes rather than a single monolith. For efficient retrieval, the dialog system needs to decide which indexes to query and even when to let the LLM to answer through its own knowledge without relying on retrieved evidence. 

Retail conversational shopping assistants need to retrieve evidence from multiple sources like reviews, product information, help pages, delivery information and more. 
These sources can include a mix of classical retrieval stores like HNSW~\cite{DBLP:journals/corr/MalkovY16} indexes built using encoder models~\cite{yu2022cocodr, DBLP:journals/corr/abs-2004-04906} or API's that link to internal or external services (eg: an API to get assembly instructions from a manufacturer's site). We also include a \texttt{no-evidence-needed} retrieval source when the LLM answers using its pre-trained knowledge. 
\begin{figure*}[ht]
    \centering
    \subfloat[Single-Step Retrieval in RAG]{\includegraphics[height=4.5cm,trim=0 10 -30 -15,]{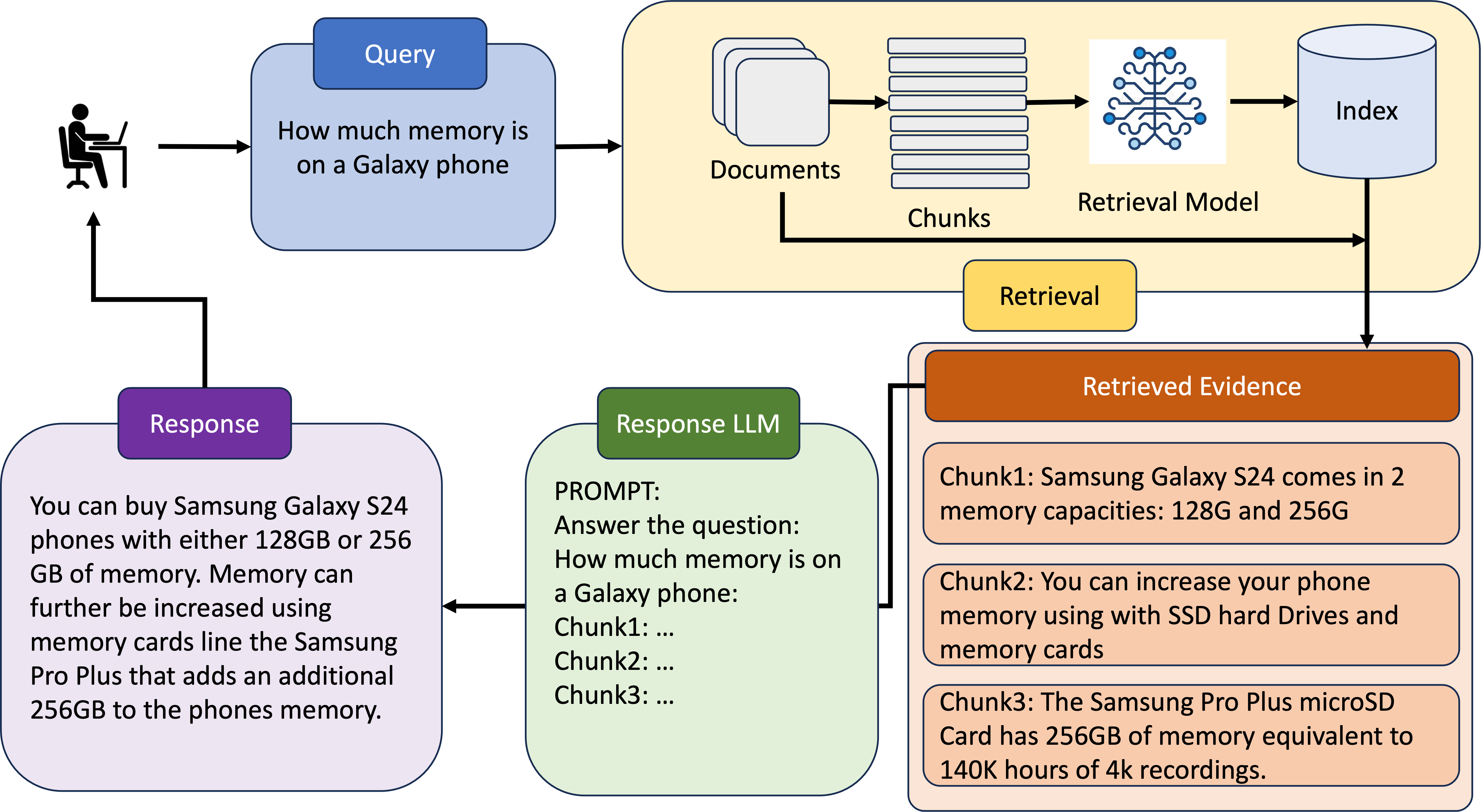}\label{fig:single-step-rag}}
    \hspace{0.2cm}
    \subfloat[Multi-Step Sequential Retrieval for RAG]{\includegraphics[height=4.5cm,trim=0 -5 -25 0,]{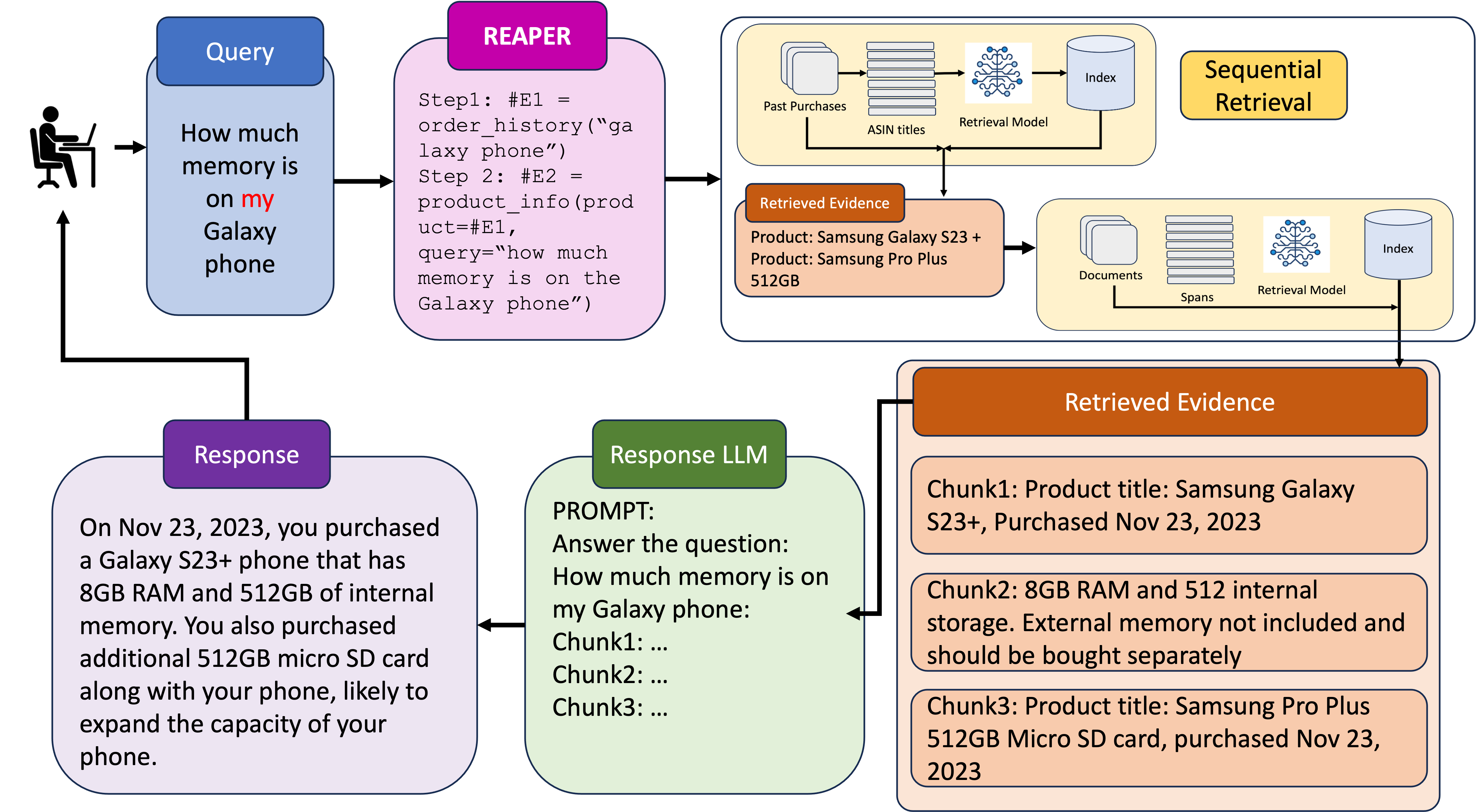}\label{fig:multi-step-RAG}}
    
    \caption{Traditional RAG systems rely on retrieving evidence in parallel from one or more sources. Conversational shopping can include features like personalization (questions about past purchases, preferences, subscriptions, etc), shopping recommendations, and more that require multi-step retrieval. These use cases can be complex enough that either an Agent is required to identify the steps, or retrieval needs its own CoT planner. We introduce REAPER for CoT retrieval planning}
    \label{fig:rufus-use-cases}
\end{figure*}

Each retrieval source or retriever is associated with latency and hardware costs. Thus, dialog systems rarely call all retrievers for every query. Instead, they need to decide which retrievers to invoke for a given query. The situation is further complicated when the retrieval itself can become multi-step. Figure~\ref{fig:rufus-use-cases} shows a scenario where changing the query from \textit{"How much memory is on \textbf{a} Galaxy phone"} to \textit{"How much memory is on \textbf{my} Galaxy phone"} significantly changes the retrieval plan. In the first case, the retrieved evidence comes from information about Galaxy phones in general. Any sufficiently large generic LLM will be able to answer this from its pre-trained knowledge. For the second question though, we first need to identify the exact phone that was purchased by the customer and give specific information pertaining to that phone. 

LLM Agents ~\cite{Shinn2023ReflexionLA, yao2022react} are able to handle the retrieval complexity by interleaving retrieval and reasoning calls. However, each reasoning step directly adds to the latency of the systems. For large models this latency cost is significant -- in the order of multiple seconds. Multi-agent systems~\cite{clarke2022one, fang2024multi} use classifiers to route the query to an appropriate Agent (\textit{question-pairing}) or use multiple Agents to generate candidate responses and a final Agent to select the best response (\textit{response-pairing}). Question-pairing gates a powerful LLM using a classifier and thus can limit the ability of the LLM. On the other hand, response-pairing adds complexity, hardware costs and latency to the system by requiring multiple LLMs to process the query in order to generate a response.


\subsection{Our Contribution}
In this paper, we introduce REAPER -- a \textbf{REA}soning based \textbf{P}lann\textbf{ER} -- for efficient retrieval required for complex queries. Using a single and much smaller LLM, REAPER generates a plan that includes the tools\footnote{Borrowing from Agent literature, we treat retrievers as tools. However, we also invoke tools that perform supplementary tasks like time conversions. Thus, tools are a superset of retrievers.} to call, the order in which they should be called and the arguments to each tool. By generating the entire retrieval plan in a single step and using a smaller LLM, we are able to minimize the latency cost as compared to single- or multi-Agent systems and still maintain the response quality. 
REAPER achieves 95\% accuracy when selecting the right tool sequence and 92\% accuracy on generating the correct tool arguments. We also show that compared to classification based question-pairing systems, REAPER is able to easily scale to new retrieval sources (tools) with very little training data and to new use cases using the current tools with just in-context examples. 

Though our architecture follows the mold of multi-Agent systems, it does not implement communication between the LLMs, which is the key element of such systems. Hence, we consider REAPER a stand-alone planner rather than a multi-Agent system. In this paper, we keep the response generation LLM constant and focus on the retrieval planning capabilities of REAPER rather than the response quality, with the understanding that with better evidence retrieval the response LLM will generate a better answer. 

\section{Literature Review}

Open Domain Question Answering is the task of accurately answering a query by \textit{retrieving} relevant documents, and interpreting them via a \textit{reader}. Extractive Readers predict an answer span from the retrieved documents~\cite{DBLP:journals/corr/abs-2004-04906, DBLP:journals/corr/abs-1911-03868}. Generative Readers generate answers in natural language using sequence-to-sequence models~\cite{DBLP:journals/corr/abs-2007-01282, DBLP:journals/corr/abs-2009-12756}. With the advent of LLMs, retrieval augmented generation (RAG) has gained popularity. \citet{gao2023retrieval} provide a survey of current RAG methods. Almost all of RAG research focuses on how retrieved evidence can be used to improve some quality metric of the generated response~\cite{izacard2022atlas, lazaridou2022internet, chan2024rq, khattab2022demonstrate}. For example,~\citet{izacard2022atlas} jointly train a retriever and LLM to improve the generation perplexity, while \citet{lazaridou2022internet} improve answer quality by generating multiple answer candidates for each retrieved evidence and the ranking them using the LLM. 

Multi-Hop QA (MHQA) requires a model to \textit{reason} over several steps and retrieved evidences to reach an answer. Similar to RAG, MHQA research focuses on how LLMs use evidence rather than how to retrieve the correct evidence~\cite{mavi2022survey}. For example, \citet{khattab2022demonstrate} and ~\citet{yao2022react} tackle complex Chain-of-Thought (CoT) reasoning by interleaving retrieval and reasoning steps in different ways. On the other hand, \citet{xu2023rewoo} introduce REWOO (Reasoning WithOut Observation), in which they argue that generating the complete plan in a single step and then executing it allows for more accurate planning. In a real-world application like conversational shopping assistants, completing the planning in a single step can help reduce the overall latency by limiting the LLM calls. 

In MHQA and other Chain-of-Thought (CoT) reasoning tasks, retrieval is the main bottleneck~\cite{mavi2022survey}. Very few prior work, though, consider the problem of efficient retrieval in RAG, MHQA or CoT systems. Even multi-Agent systems that consider retrieving the most relevant evidence, focus on answer quality instead of efficiency. \citet{fang2024multi} use an LLM to route the customer query to a one of three LLMs trained to either \textit{chit-chat}, \textit{recommend a product} or \textit{ask a question}. Multi-step retrieval, though, would still require multiple calls to the system, perhaps with the user doing part of the planning by providing more information to the clarifying question. \citet{clarke2022one} use a similar approach and use an LLM to select from responses of several Agents (\textit{response-pairing}). They also compare it to an approach of using a different classifiers to route the query to a single LLM specialized for that query shape (\textit{question-pairing}). They find that the response-pairing generates better answers but comes with significantly more complexity, while question-pairing is faster and cheaper but at the cost of lower answer quality. \citet{jeong2024adaptive} propose Adaptive-RAG, where they improve RAG time by teaching a smaller LLM to dynamically decide whether to use 1) no-evidence, 2) single-step RAG and 3) CoT with interleaved reasoning and retrieval steps. However, CoT still needs interleaving. 

REAPER combines concepts from Adaptive-RAG, question-pairing and ReWOO. We propose an architecture, where REAPER -- a smaller LLM -- generates the retrieval plan via CoT reasoning, 
and a large LLM uses the evidence to generate the appropriate response. An exemplar system diagram is shown in Figure~\ref{fig:rufus-use-cases}. 

\section{Problem Statement}
\label{sec:prob-statement}

\begin{figure}[t]
    \centering
    \includegraphics[width=0.99\linewidth]{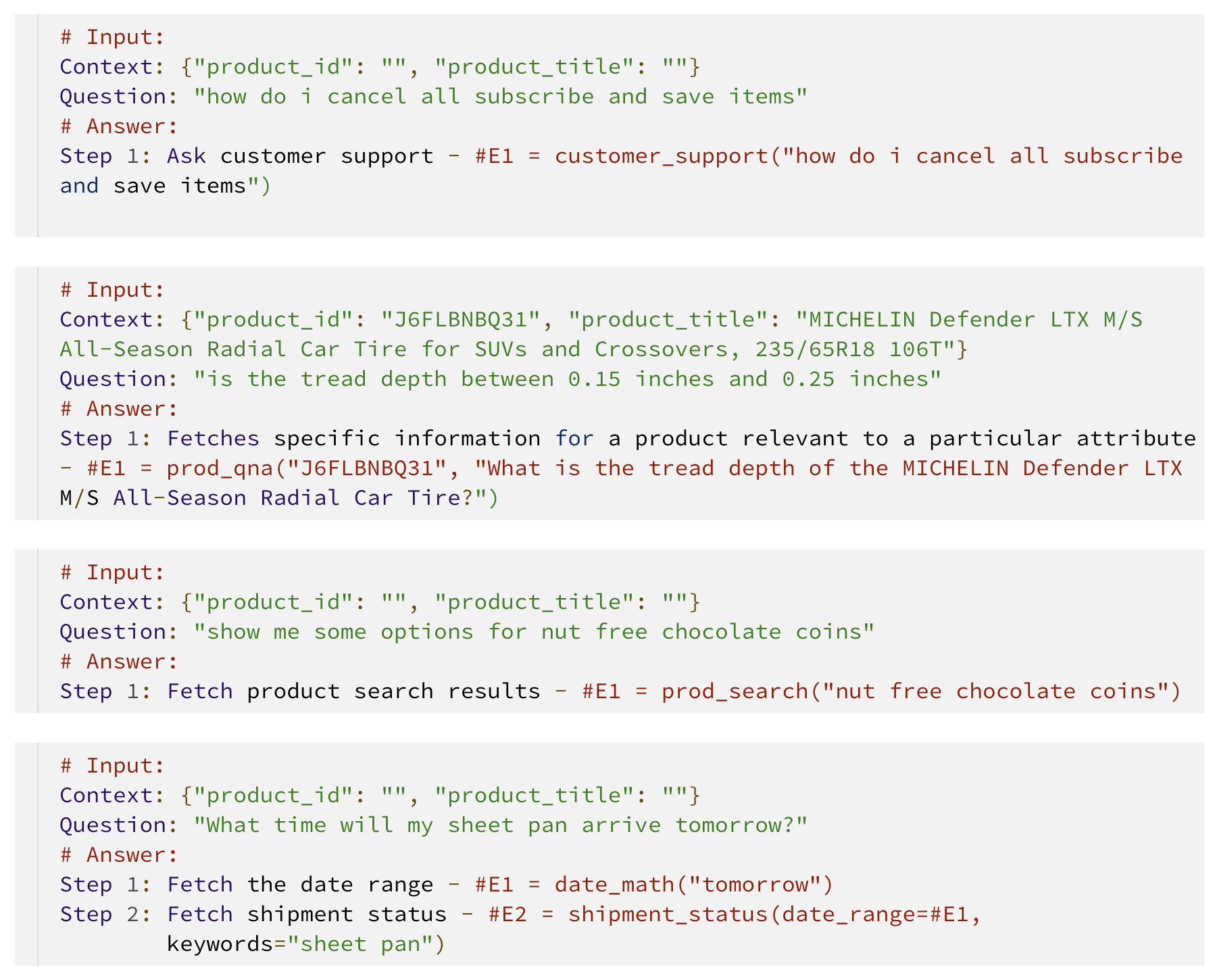}
    \caption{Example: REAPER plans. Note that REAPER is able to incorporate context (second plan) and generate multi-step retrieval plans when necessary (last plan)}
    \label{fig:reaper-plans}
\end{figure}

Our objective is to allow conversational systems to scale to queries requiring CoT retrieval plans (both single and multi-step) and new use cases without incurring the high latency and hardware cost of an Agent LLM, in a data efficient manner. We do this by moving CoT reasoning specific to retrieval to a specialized, smaller LLM. This REAsoning based PlannER (REAPER) takes as input, the customer query and contextual information. Figure~\ref{fig:rufus-use-cases}\subref{fig:multi-step-RAG} shows an example of a query requiring multi-step retrieval. For a conversational shopping assistant, a popular use-case is queries about products. Thus, when available, we provide the product information as context to REAPER. Figure~\ref{fig:reaper-plans} shows example plans where the user can ask a question with or without product context. Based on the conversational system, the context can be extended to other information like conversational history, date/time at which the question is asked, user information, url or identifier of the page on which the question is asked, etc. To generate retrieval plans, we require REAPER to:

1. Understand all the available tools used for generating evidence. 

2. Generate a retrieval plan that can work for no-evidence-retrieval, single-step retrieval and multi-step retrieval. Since REAPER will likely be the ingress point into the conversational system, the plans for all of these should be generated using the same prompt. 

3. Since REAPER mistakes can propagate all the way to the ultimate response, REAPER needs to achieve high accuracy in tool selection, sequencing and format and arguments of the tools. 

4. For latency and hardware gains, the REAPER LLM should be significantly smaller than the answer generation LLM of the conversational system. 
    
5. REAPER should be scalable to new retrievers or tools with minimal new data and training. 

6. REAPER should not hallucinate new tools for use cases it has not seen before and should be able to follow changes in the tool collection. Thus, it needs to retain good instruction following ability, although high performance on general-purpose instruction following is not required since the objective is to use it for the sole task of retrieval planning.


\subsection{Baselines}
\label{sec:curr-system}

The most common architectures for conversational systems include an Agent~\cite{10.1145/3173574.3174047, park2023choicemates} or a classification system that helps route the queries~\cite{jeong2024adaptive, clarke2022one, fang2024multi}. We thus use our implementations of such systems as baselines. We simulate a conversational Agent (or multi-Agent) by sequentially calling Claude-Sonnet ~\cite{claude-model-card} for identifying the steps in a multi-step retrieval plan. The number of retrieval-related calls is equal to the number of steps in the retrieval plan. 

As the classification-based baseline (question-paring multi-Agent system~\cite{clarke2022one}), we trained an ensemble of 2 Roberta models  
to classify queries into six classes. Our ensemble achieves better performance than a single classifier and thus, is a stronger baseline. We use a total of 150K queries to fine-tune the Roberta models. However, to add a new retriever (new class) we will need to collect tens of thousands of representative queries which is expensive and time-intensive.

Following classification, we use the Mistral 7B LLM for generating appropriate arguments for the retrievers or APIs. Complex multi-step retrieval is handled by assigning customer queries that need multi-step retrieval to a separate class which then initiates a static multi-step workflow. This means that for some such workflows we may need to call the Mistral LLM multiple times with the appropriate prompt that generates the arguments for the particular retriever. 
We also note that as the number and complexity of queries grows the classification based approach becomes cumbersome and does not scale. The number of classes with multi-step retrieval also grows combinatorially with the number of retrievers. 

We aim for REAPER to match or beat Roberta ensemble performance while also developing new capabilities like training-data efficiency, dealing with ambiguities and complex retrieval cases. 

For a fair comparison, we evaluate within the strengths of the classification models. Thus, we have limited the number of retrieval classes to a small number (six), follwoed by a call to the Mistral model to generate all arguments needed in the workflow for the class. 
Thus, REAPER has a harder task of generating the plan along with the right arguments, while the classification system simply needs to classify the queries to one of six classes. 







\section{REAPER} 

To develop an LLM that meets the requirements of Section~\ref{sec:prob-statement}, we need a reasonably small LLM with strong instruction following abilities. We use \ift ~\cite{jiang2023mistral} based on its performance on open IFT benchmarks. However, even with significant prompt tuning and in-context examples, the Mistral model was prone to hallucinations (see Figure~\ref{fig:mistral-plans-1} for examples). Hence, we fine-tune the model for our use case. In this section, we explain our methodology for designing the REAPER prompt and selecting fine-tuning data to maintain instruction following and eliminate hallucinations while learning the specialized task of retrieval tool planning. 

\begin{figure}[tb]
    \centering
    \includegraphics[width=\linewidth]{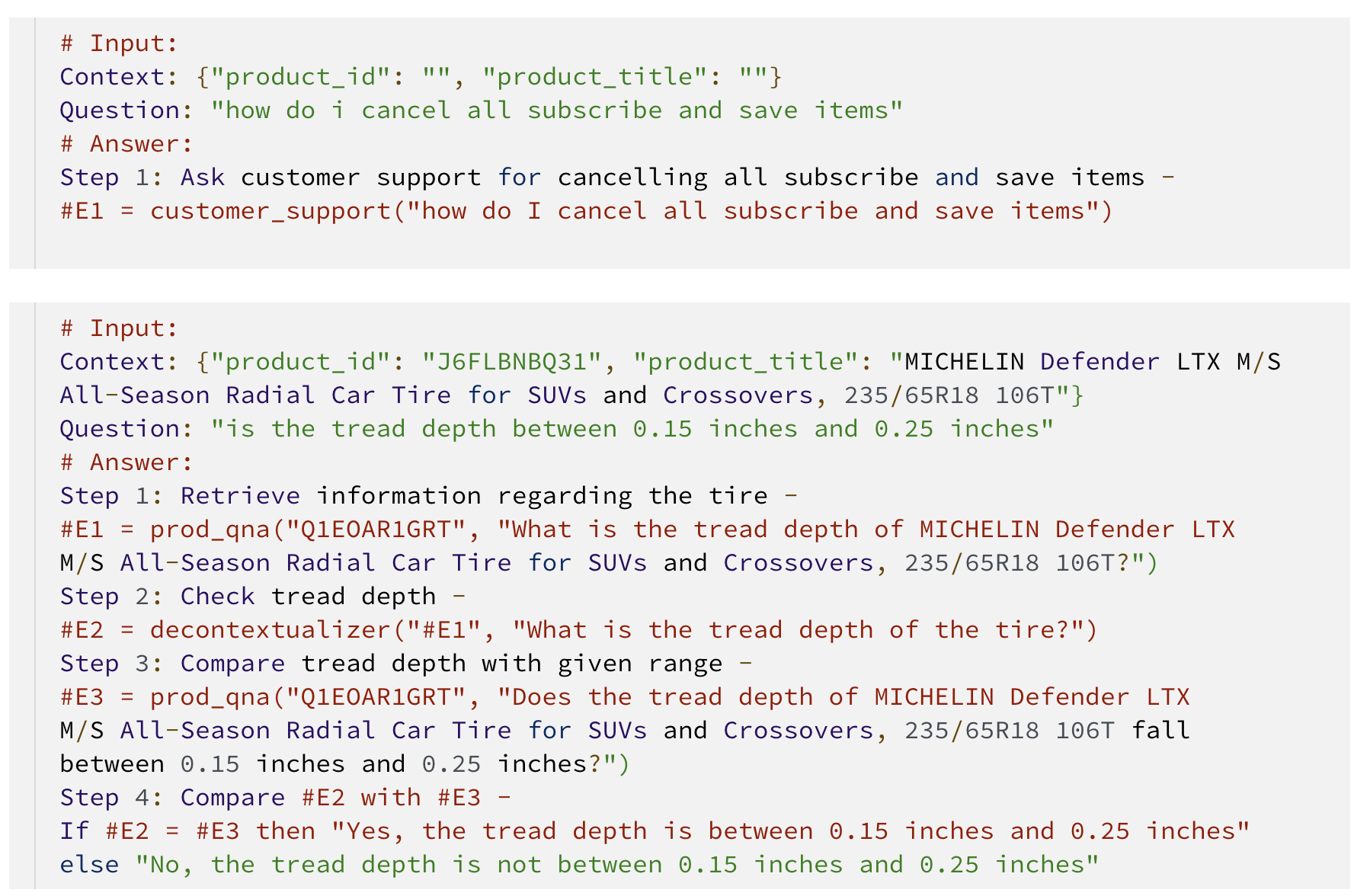}

    \centering
    \includegraphics[width=\linewidth]{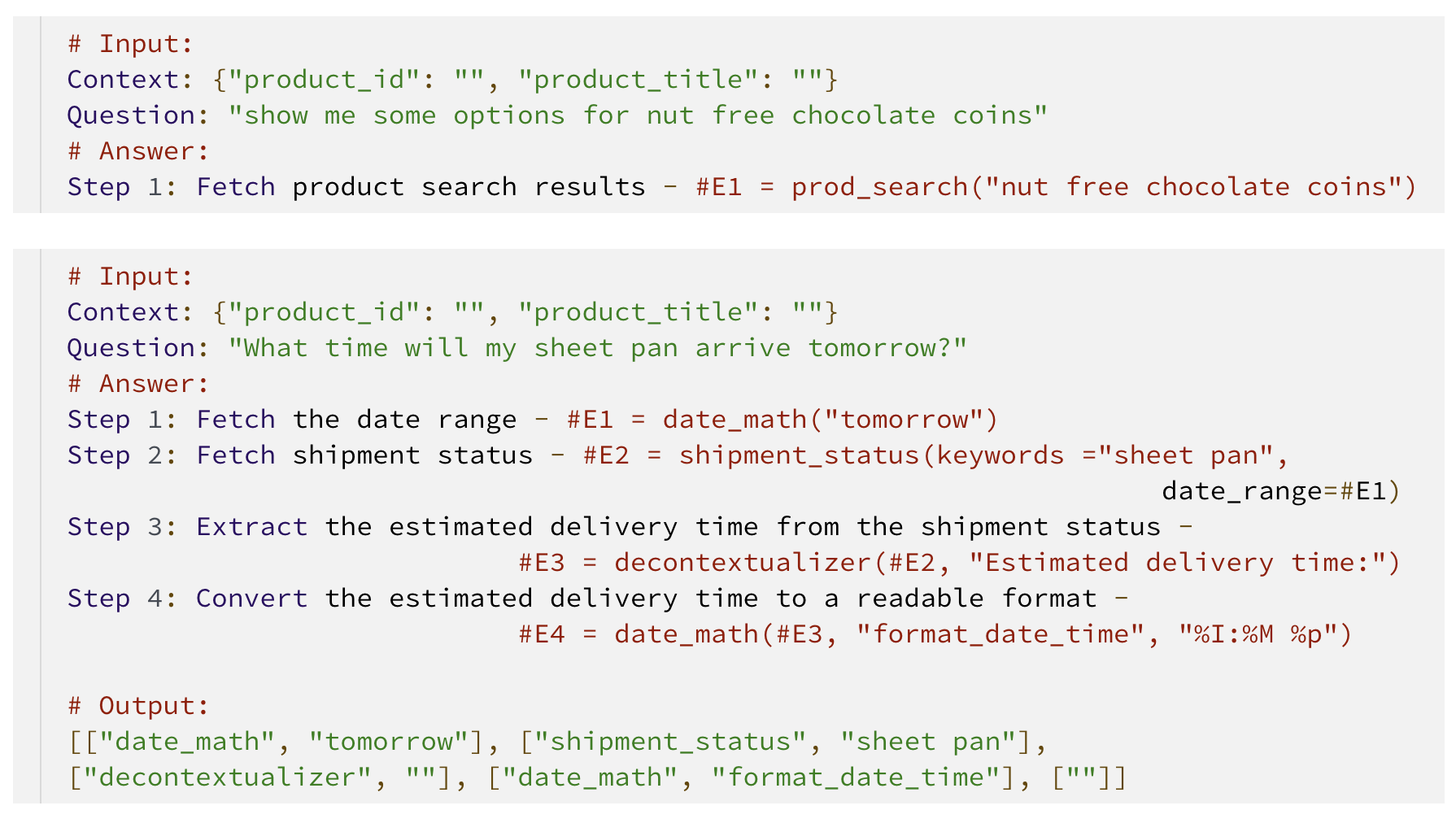}
    \caption{Example: Mistral plans. For simple queries (first and third query) Mistral produces the right plans. However, for multi-step retrieval Mistral goes beyond just retrieval evidence generation, hallucinating steps like Compare (second plan) and Extract estimated delivery time (fourth plan).}
    \label{fig:mistral-plans-1}
\end{figure}


\subsection{REAPER Prompt Design}
\label{sec:reaper_prompt}

An LLM prompt $p$ consists of an input $x$, an instruction set $I$, and a set of $m$ in-context examples, $E = {\{(\Tilde{x}_1, \Tilde{y}_1), (\Tilde{x}_2, \Tilde{y}_2), \ldots, (\Tilde{x}_m, \Tilde{y}_m)\}}$ that help the model understand the desired task. 

In our prompt, $I$ includes instructions like the role of the LLM as well as all the tools $T = \{\Tilde{f}_1, \Tilde{f}_2, \ldots, \Tilde{f}_t\}$. The tools are essentially API calls that the REAPER LLM needs to understand. We provide the tool name, tool signature, its natural language description and an example usage in the prompt. Finally, $I$ contains task instructions and constraints. Exemplar elements of the prompt are shown here: 

\begin{minted}[breaklines=true, showspaces=False, frame=lines, breaksymbol=, fontfamily=zi4, fontsize=\footnotesize]{text}

### Role:
You are an AI assistant to a salesperson at a big retail store. Your goal is to find the right information to help the salesperson answer the customer's question. 

### System Instruction:
Your goal is to generate a step by step plan using the tools listed below to get the information needed to answer a customer question. The output of one tool can be fed to another in a sequential manner. Each step may use only one tool. Some parameters of a tool can be generated with help of provided capabilites.

The set of the candidate tools, their definitions, example usages are:

1. prod_qna - Tool: Fetches specific information for a particular aspect or attribute of the product . It needs a product ID and a query as input. 
...
\end{minted}






For our case, the input $x = \{q, c\}$ includes the customer query, $q$ and page context, $c$. A customer can reference information gathered during their shopping journey. So on a product Detail Page (DP), we include the product title in $c$. On non-product pages, like Search Results Page, Landing Page, Checkout, etc., $c$ is empty. Page context is necessary for anaphoric and contextual de-referencing. For example, when a customer asks \textit{"What is your favorite color"} when say, they open the shopping app, they are simply engaging in small talk with the conversational system. However, if they ask the same question on the DP of say a t-shirt that is available in multiple colors, they expect the response in the context of that t-shirt. In former (small-talk) case, the LLM may wish to answer with no evidence along the lines of \textit{"I am an AI assistant and do not have favorite colors"} (paraphrased for brevity). In the latter case, it will use evidence from reviews to deduce the popular colors or sentiment around different colors to form an answer. 




\subsection {REAPER Data Generation Approach}

For REAPER we aim to balance two contrasting objectives: 1) the model needs to generate the plan for an in-domain task in a precise format with the right tool signatures, 2) the model should be able to understand changes in the input including nuanced changes in instructions and tools and adapt the plans accordingly. We can meet the first requirement by training the model with a large number of precise REAPER plans. However, this causes the model to overfit on the planning task and it loses its instruction following and reasoning capability. On the flip side, without enough plans in the training data, we see that the model tends to hallucinate responses. 


An important design consideration for fine-tuning an LLM for instruction following is to provide it with a diverse set of prompt and output pairs so that it does not overfit on a specific task template and catastrophically forget its ability to closely follow instructions. This becomes particularly challenging since our primary task is retrieval planning and so the scope of 
introducing diversity in the instruction set is limited. We develop the following modules for different aspects of input and output data diversity. 

\subsubsection{Tool Evolve (TEvo): Evolution of Base Tool Prompt}We introduce a novel module  \textbf{Tool Evolve (TEvo)} that takes the tool prompt as input and produces a semantically similar prompt in a way that the output, $y$ does not change. The technique is similar to introducing adversarial noise into images for building robust image classifiers \cite{10.1145/3485133}. To force our model to pay attention to tools and their corresponding descriptions, we select the tool(s) required to for a particular query and a random subset of the remaining tools to include in the instruction section of our prompt. We also create a pool of name variation and description paraphrases for each tool and sample from these. For example, to answer questions about a product, our tool names can be \texttt{prod\_qna}, \texttt{product\_information}, \texttt{product\_facts}, etc. Finally, we also vary in-context examples by sampling from a small pool of human generated plans.



\subsubsection{Tool-Task Generator (TTG)}
\label{ssec:ttg}
In Wizardlm~\cite{xu2023wizardlm}, the authors increase the complexity of simple tasks and add these tasks to the IFT training data in order to improve the instruction following capability of an LLM.  Similarly, we introduce an approach to create diversified tasks related to retrieval planning. This forces the LLM to understand tools and retrieval plans. 


Given a primary task of generating a retrieval plan $T_{prim} = (\mathbf{x}, \mathbf{y})$, where $\mathbf{x}$ is the input containing the query $q$ and context $c$, we transform the task into multiple related tasks. 



\begin{equation}
    T_i = f_i(T_\text{primary})
\end{equation}

where $f_i$ is the task-specific transformation function. The secondary tasks are target specific capabilities like: 

\begin{enumerate}
    \item Generate query from a plan: $T_1 = \text{generate\_query}(\mathbf{y})$, 
    \item Complete partial plan: $T_2 = \text{complete\_partial\_plan}(\mathbf{x}, \mathbf{y}_\text{partial})$, 
    \item Identify the right tools: $T_3 = \text{identify\_tools}(\mathbf{x})$, 
    \item Masked step completion: 
    
    $T_4 = \text{complete\_masked\_step}(\mathbf{x}, \mathbf{y}_\text{masked})$, 
    \item Reordering steps in coorect sequence: 
    
    $T_5 = \text{reorder\_steps}(\mathbf{x}, \mathbf{y}_\text{shuffled})$, 
    \item Masked parameter identification: 
    
    $T_6 = \text{identify\_masked\_params}(\mathbf{x}, \mathbf{y}_\text{masked})$, 
    \item Limit tools in plan to only the ones specified in the prompt: 
    
    $T_7 = \text{use\_provided\_tools}(\mathbf{x}, \text{tools})$ 
\end{enumerate}
The TTG module creates diverse task-target pairs $\{(\mathbf{p}_1, \mathbf{y}_1), 
\dots, \newline(\mathbf{p}_N, \mathbf{y}_N)\}$ by applying the secondary task transformations $f_i$ to the primary task $T_\text{primary}$. We then sample from this diverse set of tasks to fine-tune the REAPER model, with the aim of developing a robust understanding of the overall task structure and enhancing its ability to generate retrieval plans for complex customer queries.

\subsubsection{Diverse Query Sampler (DQS)}
In addition to adding diversity in prompt using TEvo and diversity in output using TTG, we also diversify the input $x_k$, which is a question from a customer with a page context. Similar queries cause the model to fixate on particular query shapes leading to performance degradation when the model encounters out-of-distribution cases. To this end, we propose \textbf{Diverse Query Sampler (DQS)} that automatically generates a sample of customer queries that are semantically dissimilar. We annotate the diverse samples with relevant tools and parameters based on the conversational context. 

Given a high-quality curated initial (small) pool of customer queries, $\mathbf{Q}_\text{initial} = \{\mathbf{x}_1, \mathbf{x}_2, ..., \mathbf{x}_N\}$ and a larger pool of generated or sampled customer queries $\mathbf{Q}_\text{large}$, DQS introduces diversity by:

1. Generating BERT-based embeddings $\mathbf{e}_i$ for each query $\mathbf{x}_i$ in $\mathbf{Q}_\text{large}$, where $\mathbf{e}_i = \text{RoBERTa}(\mathbf{x}_i)$.

2. Calculating the pairwise cosine similarity between the query embeddings in $\mathbf{Q}_\text{initial}$ and $\mathbf{Q}_\text{large}$ to obtain a similarity matrix $\mathbf{S}$, where $\mathbf{S}[i,j] = \cos(\mathbf{e}_i, \mathbf{e}_j)$ for $\mathbf{x}_i \in \mathbf{Q}_\text{initial}$ and $\mathbf{x}_j \in \mathbf{Q}_\text{large}$.

3. Identifying the most similar and most dissimilar pairs of queries in $\mathbf{Q}_\text{large}$ based on the similarity matrix $\mathbf{S}$. Let these be the set of queries at the extremes of diversity, denoted as $\mathbf{Q}_\text{extreme}$.

4. Removing $\mathbf{Q}_\text{extreme}$ from $\mathbf{Q}_\text{large}$ to obtain a refined pool $\mathbf{Q}_\text{refined} = \mathbf{Q}_\text{large} \setminus \mathbf{Q}_\text{extreme}$.

5. Randomly sampling a subset of queries from the larger pool $\mathbf{Q}_\text{large}$ to obtain the final diverse set of customer queries $\mathbf{Q}_\text{diverse}$. The size of $\mathbf{Q}_\text{diverse}$ is chosen such that $|\mathbf{Q}_\text{diverse}| = |\mathbf{Q}_\text{initial}|$.

The goal of this process is to ensure that the queries in $\mathbf{Q}_\text{diverse}$ have a balanced representation of semantic diversity, reducing model bias and enhancing the ability of REAPER to generalize across different customer information needs.

\subsubsection{General purpose IFT data}
To retain the model's instruction following ability, we include general purpose instruction fine-tuning datasets in addition to the REAPER tool planning data. We use ShareClaude and open-source tool usage data from ToolAlpaca \cite{tang2023toolalpaca} for generic IFT data. We call this dataset \nileift\ . To further enhance the model's ability to follow nuanced changes in the input instruction, we utilize a framework inspired from Evol-Instruct \cite{xu2023wizardlm} that automatically generates more complex IFT data by adding constraints on simple instructions and samples to maintain a roughly equal proportion of query complexity as measured by the prompt length used for the task. We call this dataset \nileevolve. Our final fine-tuning dataset, \reaperift\ is a combination \nileift/\nileevolve\ and our tool-annotated queries which are diversified using TEvo, TTG and DQS. 




\section{Experiments}

In this section we present experimental setting and results  comparing REAPER against \ift, Claude3-Sonnet and the our ensemble classifier described in Section~\ref{sec:curr-system}. Our experiments show the need for fine-tuning for the plan generation as larger models like Claude3-Sonnet or Mistral tend to hallucinate. We also present ablation studies to show the impact of different components that we proposed to create our training data. 

\begin{figure}[t]
    \centering
    \includegraphics[width=\linewidth]{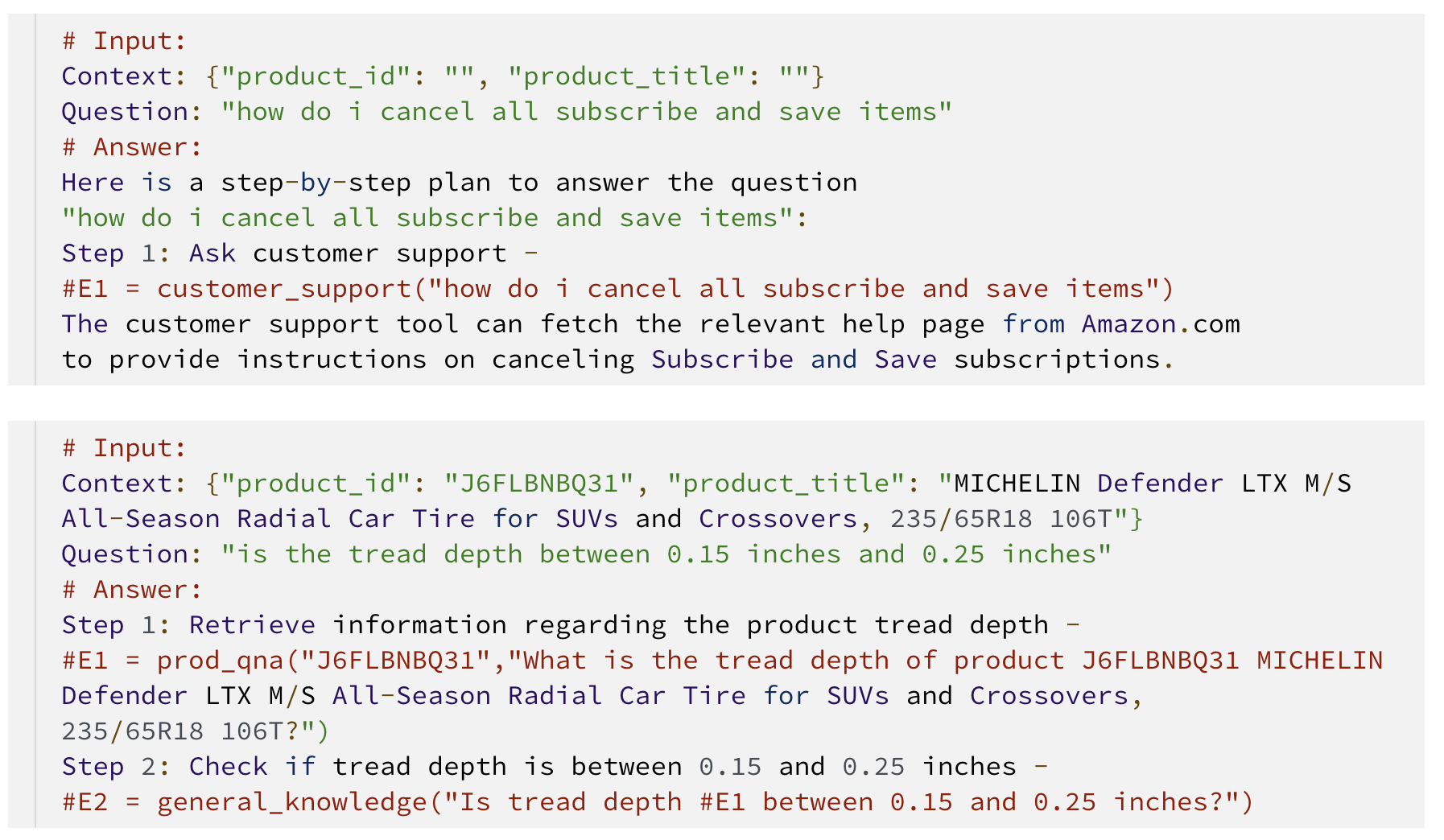}

    \centering
    \includegraphics[width=\linewidth]{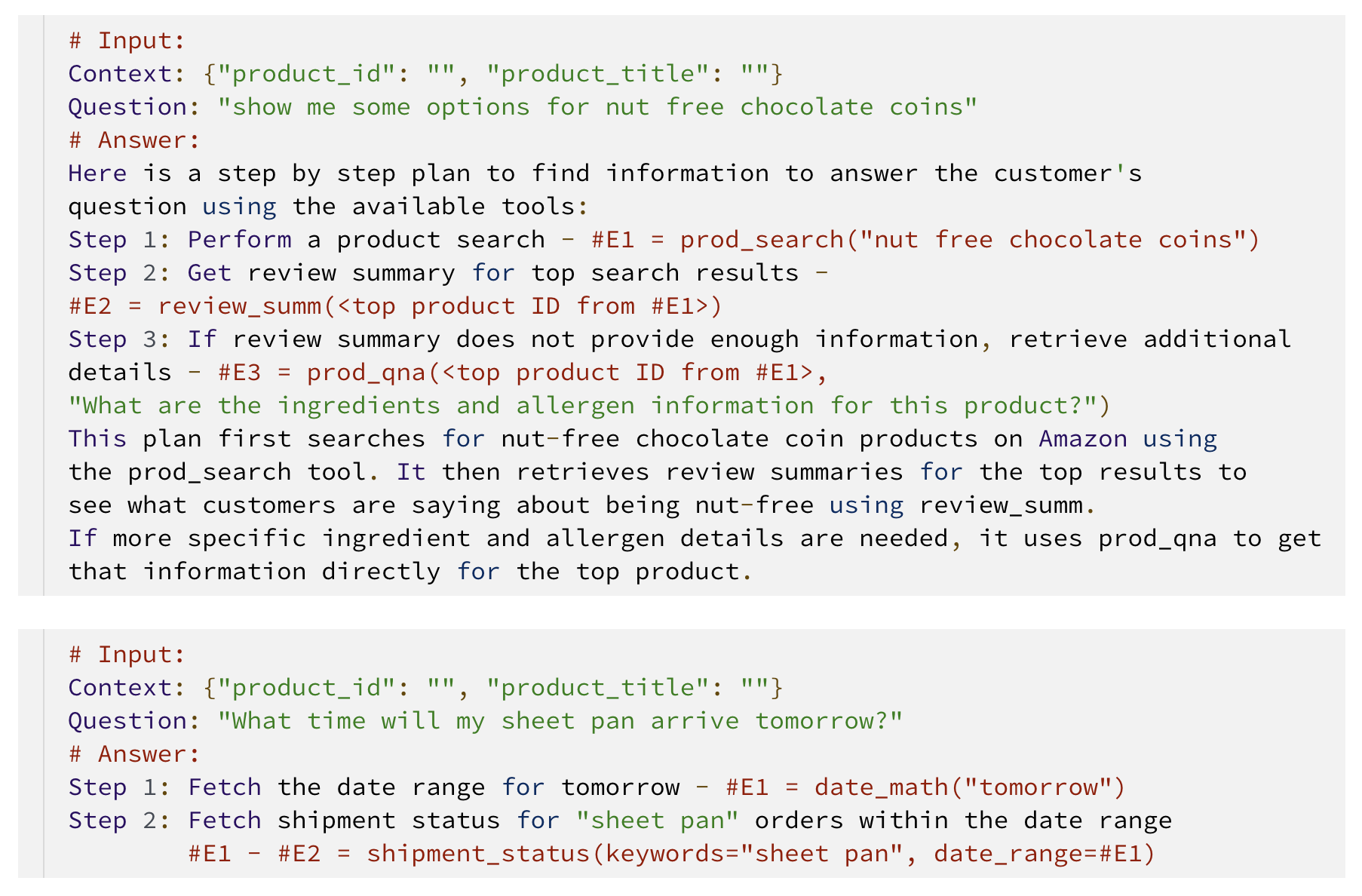}
    \caption{Example Claude3-Sonnet plans. Similar to Mistral, we also see hallucinations in plans. For example, extra step in the second plan and extra explanation in the third plan. With more effort it may be possible to reduce hallucinations. Claude3-Sonnet latency though, is 10x of Mistral latency.}
    \label{fig:claude-plans-2}
\end{figure}

\subsection{Comparison with Open Models}
It is generally desirable to solve a task using a language model using in-context learning. Despite several weeks worth of effort, we could not prompt-tune \ift to reach the target performance. Figure~\ref{fig:mistral-plans-1} shows examples of the plans generated by \ift with prompt tuning. We also tested out our REAPER prompt on Claude3-Sonnet (Figure~\ref{fig:claude-plans-2}) and could not prevent hallucinations. In addition, Claude3-Sonnet latency was \textbf{$\sim$2s} per step as compared to \textbf{207ms} for the entire plan for REAPER and Mistral models -- introducing an order of magnitude latency cost in addition to more powerful hardware required to run Sonnet. 

\begin{table*}[tb]
    
\caption{Comparison of REAPER performance with Baselines on six retrieval classes}
\label{tab:overall_performance}
\begin{tabular}{|c|lll|llll|llll|}
\hline
                               & \multicolumn{3}{c|}{\begin{tabular}[c]{@{}l@{}}Mistral\\ (No fine-tuning)\end{tabular}} & \multicolumn{4}{c|}{Ensemble Classifier}                                                                                                   & \multicolumn{4}{c|}{REAPER}                                                                                                                   \\ \hline
Classes                        & \multicolumn{1}{l|}{P}             & \multicolumn{1}{l|}{R}             & F1            & \multicolumn{1}{l|}{P}  & \multicolumn{1}{l|}{R}   & \multicolumn{1}{l|}{F1} & \begin{tabular}[c]{@{}l@{}}\#Training\\ Examples\end{tabular} & \multicolumn{1}{l|}{P}   & \multicolumn{1}{l|}{R}   & \multicolumn{1}{l|}{F1} & \begin{tabular}[c]{@{}l@{}}\#Training\\ Examples\end{tabular} \\ \hline
Customer Support                     & \multicolumn{1}{l|}{90}            & \multicolumn{1}{l|}{61}            & 73            & \multicolumn{1}{l|}{95} & \multicolumn{1}{l|}{81}  & \multicolumn{1}{l|}{88} & \multicolumn{1}{c|}{24621}                                                              & \multicolumn{1}{l|}{95}  & \multicolumn{1}{l|}{94}  & \multicolumn{1}{l|}{94} &   \multicolumn{1}{c|}{1127}                                   \\ \hline
Shipment Status                   & \multicolumn{1}{l|}{97}            & \multicolumn{1}{l|}{72}            & 83            & \multicolumn{1}{l|}{96} & \multicolumn{1}{l|}{96}  & \multicolumn{1}{l|}{96} & \multicolumn{1}{c|}{16150}                                                              & \multicolumn{1}{l|}{98}  & \multicolumn{1}{l|}{94}  & \multicolumn{1}{l|}{96} & \multicolumn{1}{c|}{996}                                                              \\ \hline
Product Search                 & \multicolumn{1}{l|}{82}            & \multicolumn{1}{l|}{65}            & 72            & \multicolumn{1}{l|}{84} & \multicolumn{1}{l|}{99} & \multicolumn{1}{l|}{91} & \multicolumn{1}{c|}{38683}                                                              & \multicolumn{1}{l|}{91}  & \multicolumn{1}{l|}{100} & \multicolumn{1}{l|}{95} & \multicolumn{1}{c|}{1289}                                                              \\ \hline
Product QnA & \multicolumn{1}{l|}{47}            & \multicolumn{1}{l|}{80}            & 59            & \multicolumn{1}{l|}{98} & \multicolumn{1}{l|}{97}  & \multicolumn{1}{l|}{98} &                 \multicolumn{1}{c|}{30813}                                              & \multicolumn{1}{l|}{93}  & \multicolumn{1}{l|}{99}  & \multicolumn{1}{l|}{96} & \multicolumn{1}{c|}{1045}                                                              \\ \hline
Review Summary                 & \multicolumn{1}{l|}{79}            & \multicolumn{1}{l|}{93}            & 85            & \multicolumn{1}{l|}{99} & \multicolumn{1}{l|}{96}  & \multicolumn{1}{l|}{97} & \multicolumn{1}{c|}{9875}                                                              & \multicolumn{1}{l|}{100} & \multicolumn{1}{l|}{94}  & \multicolumn{1}{l|}{97} & \multicolumn{1}{c|}{594}                                                              \\ \hline
No-retrieval             & \multicolumn{1}{l|}{85}            & \multicolumn{1}{l|}{67}            & 75            & \multicolumn{1}{l|}{98} & \multicolumn{1}{l|}{99}  & \multicolumn{1}{l|}{99} & \multicolumn{1}{c|}{35934}                                                              & \multicolumn{1}{l|}{100} & \multicolumn{1}{l|}{93}  & \multicolumn{1}{l|}{96} & \multicolumn{1}{c|}{1245}\\
\hline 
\hline
Tool Accuracy        & \multicolumn{3}{c|}{72\%}  & \multicolumn{4}{c|}{94\%}    & \multicolumn{4}{c|}{\textbf{96\%}} \\
\hline 
\hline
Argument Accuracy        & \multicolumn{3}{c|}{--}  & \multicolumn{4}{c|}{{88\%}}    & \multicolumn{4}{c|}{\textbf{92\%}}


\\ \hline
\end{tabular}
\end{table*}

\subsection{Comparison with Classifier-based Planners}

Our ensemble classifier is described in Section~\ref{sec:curr-system}. We evaluate the models along two dimensions:
\begin{enumerate}[leftmargin=0.1cm]
    \item \textbf{Tool Selection:} Given a query, we manually evaluate if the model selects the correct tool(s) in the proper sequence to retrieve evidence. As multi-step retrieval is just another class in the question-pairing system, this metric can be directly compared to the classification metrics in question-pairing.  We present accuracy, precision, recall and F1-score for this evaluation. 
    \item \textbf{Argument Extraction:} The other aspect of REAPER is the accuracy of the arguments fed to the tools (refer Figure~\ref{fig:reaper-plans}). In the current setting, only two tools \texttt{prod\_search} and \texttt{shipment\_status} require the arguments different from the customer query. We thus restrict our evaluation to plans involving these tools. Our baseline system uses the \ift model to generate the arguments using a prompt specifically tuned for each class. REAPER does both tool selection and argument generation for all the tools in a single prompt. 
\end{enumerate}

\vspace{0.3cm}
Table~\ref{tab:overall_performance} shows the training data size and precision, recall and F1 metrics for REAPER, the baseline ensemble and \ift\ with only in-context tuning. Given Mistral tool accuracy is so low, we did not compute the Argument accuracy for it. 

\subsubsection{Evaluation Datasets}

Since conversational shopping is new, traffic distributions are still skewed towards existing traffic patterns instead of the new conversational use cases. So instead of sampling traffic to generate our evaluation set, 
we use a balanced evaluation set of 600 queries such that the corresponding plans have a roughly equal proportion of tools (classes) 

\subsection{Ablation Study of Data Components}

We show the effectiveness of TEvo, DQS and IFT-Evolve in Table \ref{tab:reaper_ablation_study}. Apart from tool selection accuracy, we also investigate how these components help the model to follow instructions by introducing an adversarial prompt where we remove the \texttt{prod\_qna} tool and obtain REAPER predictions on the test dataset. Since the model is instructed to use only the specified tools it should not use \texttt{prod\_qna} to generate a plan. 
We measure instruction following capability \footnote{ We plan to evaluate performance on open IFT benchmarks, though we expect the performance to be similar to our domain specific evaluation} as (1-proportion of plans that use \texttt{prod\_qna}). 

\begin{table}[b]
\centering
\caption{Impact of training data selection components}
\label{tab:reaper_ablation_study}
\scriptsize
\begin{tabular}{@{}lrr@{}} 
\toprule
Model                    & Tool Accuracy   & Instruction Following \\ 
\midrule
REAPER       & 0.96  & 1.00\\
REAPER w/o \nileevolve    & 0.95 &  0.76\\
REAPER w/o \nileevolve\space and DQS   &     0.87 & 1.00\\
REAPER w/o \nileevolve, TEvo and DQS         &  0.91 & 0.85\\
\bottomrule
\end{tabular}
\end{table}

We observe that REAPER including all three components achieves both the highest tool selection accuracy and the best instruction following as none of its plans in the adversarial setting uses the omitted \texttt{prod\_qna} tool. 
When we remove \nileevolve\ data from training and use \nileift\ instead, tool selection accuracy drops. The model also loses instruction following capability as it uses \texttt{prod\_qna} in 24\% of plans. We observe the same phenomenon when we remove the TEvo (task diversity) and DQS (input query diversity) components from training, with a marked drop in accuracy when we drop DQS. This suggests that TEvo (task diversity) is needed for instruction following capabilities and DQS is needed for the model to understand the different query shapes. 

\subsection{Effect of Training Data Proportions }

Table \ref{tab:data_proportion_study} shows the different data proportions we experimented with in the training set.
We found that it is essential to either add ShareClaude (\nileift\ ) or its evolved version -- \nileevolve\ with REAPER plans in the training set. Without the generic data the model severely overfits to the REAPER use case and catastrophically forgets its instruction following capability. In this case, we see 77\% of queries hallucinate tools that the model has seen in training but are not in the instruction (IFT score = 0.23). 
On the flip side, with only generic-IFT training data, the model achieves only 20\% accuracy for the in-domain task of tool selection. 

The proportion of the generic IFT training data does affect the performance of the model. In the middle three rows of Table~\ref{tab:data_proportion_study}, we varied the proportion of the  \nileevolve\ data from 40\% to 60\%, while the last three rows show variation in the number of REAPER planning tasks (cf. Section~\ref{ssec:ttg}) per query. We see improvement in both metrics as we increase the amount of IFT data and REAPER plans up to a point, beyond which we see a drop in tool selection accuracy. We found the best balance between accuracy and instruction following is seen in a roughly 1:6 proportion of REAPER tasks and generic IFT (\nileevolve \space (100\%) + TTG (4-task)). However, we expect to tune these hyperparameters as we expand the system to new retrievers and use cases.




\begin{table}[tb]
\centering
\caption{REAPER Data Proportion Study. Note that the REAPER tasks are generated using 6K unique in-domain queries, while the Generic IFT data is from open datasets.}
\label{tab:data_proportion_study}
\scriptsize
\begin{tabular}{lcccc} 
\toprule
Training data mix & Tool & \#REAPER & \#Generic & Instruction \\
 & Accuracy & tasks &  IFT & following \\
\midrule
\nileevolve \space (0\%)+ TTG (3-task) & 0.92 & 19k & 0k & 0.23\\
\nileevolve \space (100\%)+ TTG (0-task) & 0.2  & 0k & 147k & 0.71\\
\midrule
\nileevolve \space (40\%)+ TTG (3-task) & 0.93  & 19k & 60k & 1.00\\
\nileevolve \space (50\%)+ TTG (3-task) & 0.93 & 19k & 75k & 1.00\\
\nileevolve \space (60\%) + TTG (3-task)& 0.9  & 19k & 90k & 1.00\\
\midrule
\nileevolve \space (100\%) + TTG (3-task) & 0.86  & 19k & 147k & 1.00\\
\nileevolve \space (100\%) + TTG (4-task) & 0.95  & 25k & 147k & 1.00\\
\nileevolve \space (100\%) + TTG (5-task) & 0.94  & 32k & 147k & 1.00\\

\bottomrule
\end{tabular}
\end{table}

\subsection{Analysis on Generalization}

\begin{figure}[tb]
    \centering
    \includegraphics[width=0.99\linewidth]{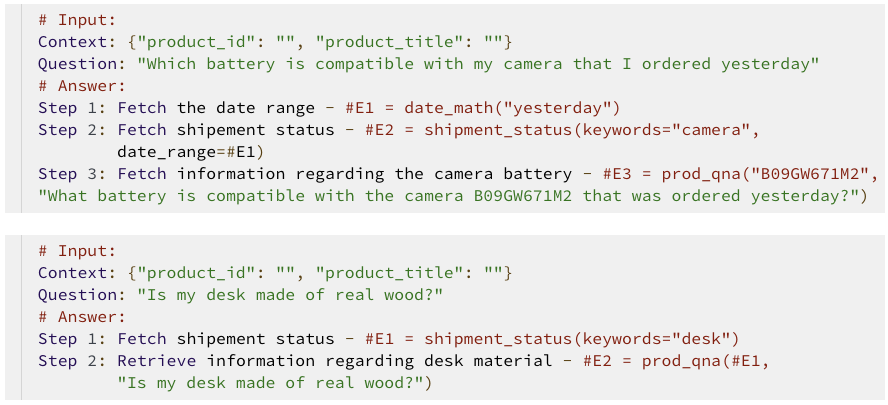}
    \caption{Plans for new use cases. The first plan hallucinates the product\_id but the second plan shows no hallucinations after we add just 25 post-purchase examples to training set}\label{fig:novel_multi_step}
\end{figure}

\paragraph{Can REAPER generate a novel plan?} Our training data has multi-step plans only in the order status context. Even so, using in-context examples alone REAPER is able to generate the right tools sequences for use-cases for query shapes it has never seen before. We did see occasional hallucinations. We added only 25 more diverse multi-step examples and REAPER was able to generate accurate brand new plans without hallucinations. Figure ~\ref{fig:novel_multi_step} shows some examples. 


\paragraph{Can REAPER be fine-tuned with limited data?}

A practical use-case of REAPER is adapting to a new retrieval source with limited amount of data. We tested this, by adding a new \texttt{no-evidence} class called \texttt{human\_small\_talk}. We added only 286 examples of this class to our training data and saw that the model achieves an F1 score of 0.92 which comparable performance to other sources trained with order of magnitude more data. 


\begin{figure}[tb]
    \centering
    \includegraphics[width=0.9\linewidth]{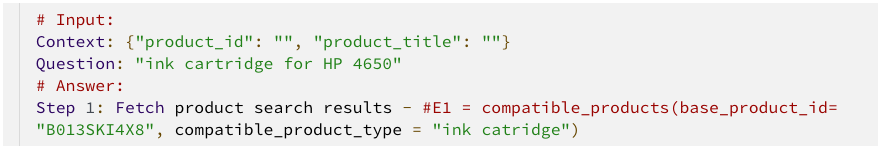}
    \caption{REAPER plans with tools not in training data}
     \label{fig:novel_tool}
\end{figure}

\paragraph{Can REAPER use a novel tool?}

We explore how well REAPER can learn to use a new tool without any training data at all and based solely on in-context examples for the tool. To do this, we added a new toolcalled \texttt{compatible\_products} with its description and a few demonstrative examples in the prompt and found that REAPER is able to generate a valid plan. For the query \textit{ink cartridge for HP 4650}, REAPER generated the plan in Figure~\ref{fig:novel_tool}. This also shows that REAPER has maintained its instruction following ability.






\section{Conclusion}

In this paper, we present REAPER -- a reasoning-based planner -- to generate retrieval plans to support RAG-based conversational systems. The planner is based on an instruction-tuned LLM and utilizes a novel data generation module to optimize the model for the specialized task of retrieval planning while still retaining the ability to follow instructions in order to scale to new use-cases. Extensive experiments show that our model is 1) data-efficient compared to typical classification models -- REAPER is trained on 6K in-domain queries, while classification models needed 150K and 2) easily scalable to new retrieval sources -- we were able to add a new retrieval sources by increasing the training data by 286 in-domain queries. It also is an order of magnitude faster (207ms as compared to 2s) as compared to Agent based systems. Finally, we observe promising results indicating our model's capability to generalize to new tools and plan structures without explicitly being trained for it. 



\bibliographystyle{ACM-Reference-Format}

\bibliography{references}

\end{document}